\documentclass[arxiv,usenatbib]{iupartex}
\usepackage{newtxtext,newtxmath}
\usepackage[T1]{fontenc}
\usepackage{ae,aecompl}

\usepackage{bm}		    
\usepackage{pdflscape}	

\newcommand{\qq}[1]{``#1''}

\newcommand{\nn}[1]{\text{#1}}

\newcommand{\B}{\bfseries}

\chardef\us=`\_

\providecommand{\noopsort}[1]{}

\title[A New Approach to Calculate Coronal Electron Density]{A New Approach to Calculate Coronal Electron Density: Simplified van de Hulst's Method}

\author[H. \c{C}akmak]{%
H. \c{C}akmak$^{1\cc}$\orcid{0000-0002-1959-6049}
\affsep \\
$^1$\.Istanbul University, Faculty of Science, Department of Astronomy and Space Sciences, 34116, Beyaz\i t, \.Istanbul, T\"urkiye
}

\corres{H. \c{C}akmak}{hcakmak@istanbul.edu.tr}

\pubyear{2023}

\doiheader{XXXXXXX/PAR.20XX.00000}
\date{
	\pSubmit{00.00.2023} 
	\pRevReq{00.00.2023}
	\pLastRevRec{00.00.2023}
	\pAccept{00.00.2023}
	\pPubOnl{00.00.0000}
}
\volume{0}
\volnumber{1}

\begin{document}
\label{firstpage}
\pagerange{\pageref*{firstpage}--\pageref*{lastpage}}
\maketitle

\begin{abstract}
Determining the electron density is a challenging task in solar corona studies, as it requires certain assumptions to be made, such as symmetric, homogeneous and radial distribution, thermal equilibrium, etc. In such studies, the observed $K$ corona brightness is based on the coronal electron density. An important paper on the calculation of electron density was published in 1950 by van de Hulst in an article titled \qq{The Electron Density of the Solar Corona}. The author developed a method with some assumptions to calculate the electron density from the observed $K$ corona brightness. We presented here, a new simplified calculation method for the coronal electron density is presented. The integral equation solution given by van de Hulst is interpreted from a different perspective and the $K$ coronal electron density is calculated using only observational data without making any additional adjustments such as successive approximations and multiple attempts.
\end{abstract}

\begin{keywords}
Sun: corona -- scattering -- polarization, Astrometry and celestial mechanics: eclipses
\end{keywords}


\section{Introduction}\label{intro}
\begin{figure*}[t!]   
	\centerline{
		\includegraphics[width=0.485\linewidth]{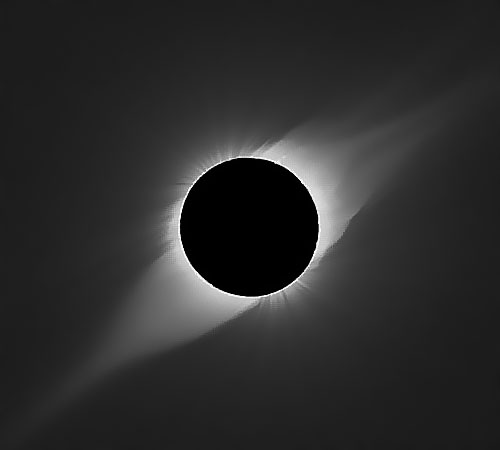}
		\includegraphics[width=0.485\linewidth]{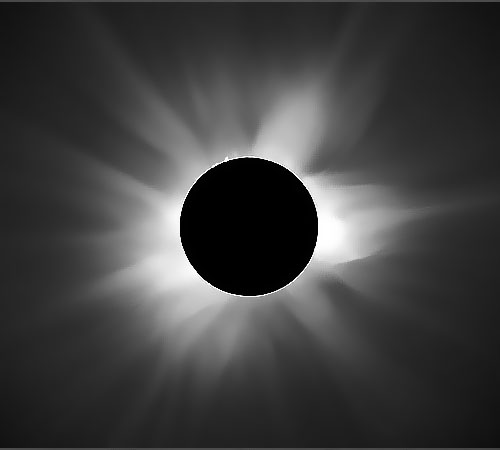}
	}
	\caption{Appearance of the solar corona according to solar cycle phase. \textit{Left}: minimum phase (on 4 October 1995), \textit{Right}: maximum phase (on 21 June 2001).} \label{corona_type}
\end{figure*}

Theoretical studies on the solar corona began with a published article by \cite{SA1879}. This work investigated the brightness and polarisation of the solar corona with regards to various particle distributions within the corona. The majority of the fundamental mathematical issues were resolved with the explanations provided here. According to this, the corona light is the composite of all the light scattered by the free electrons in the line of sight direction. The polarisation of the corona light results from this phenomenon. \cite{MM1930} further developed Schuster's theory by taking into account the limb darkening effect of the observed solar disc. Additionally, the equations for the relation between electron density and brightness were provided. \cite{BS1937, BS1938} introduced the first general formula for the electron density of the solar corona from photometric observations as a function of the solar radius;
\begin{equation}\label{eqn:1}
	N(r)=10^8 \Bigg( \frac{0.036}{r^{1.5}}+\frac{1.55}{r^6}+\frac{2.99}{r^{16}} \Bigg)
\end{equation}
where $N$ is the electron density in cm$^3$, and $r$ is the distance from the solar disc expressed in solar radius. Subsequently, corona light intensity was analysed by \cite{ACW1946} and \cite{VDH1950} according to the minimum and maximum phases of the solar cycle. They provided two distinct corona models. The type of corona during cycle maximum exhibits nearly spherical brightness distribution, and most coronal structures show a symmetric arrangement across the solar disc (see Figure \ref{corona_type}, right panel). In contrast, the type of corona during cycle minimum exhibits a concentration of coronal structures in the equatorial and polar regions (Figure \ref{corona_type}, left panel) and features with asymmetric brightness distribution. Furthermore, \cite{SK1970} developed an empirical function of the electron density which also depends on the heliographic latitude as follows;
\begin{equation}
	\begin{split}\label{eqn:2}
		N_\nn{e}(r, \phiup) &= \frac{3.09\times 10^8}{r^{16}} \big( 1-0.5\ \sin \phiup \big)\\
		&+ \frac{1.58\times 10^8}{\mathit{r}^6} \big( 1-0.95\ \sin \phiup \big)\\
		&+ \frac{0.0251\times 10^8}{r^{2.5}} \big( 1-1.0\ \sin^{0.5} \phiup \big)
	\end{split}
\end{equation}
where $N_{\rm e}$ is the electron density in cm$^3$ and $\phi$ is the heliographic latitude. The equation allows us to compute the electron density asymmetrically across the solar disc. It is thus possible to use this equation to represent changes in coronal brightness based on corona type during solar minimum or maximum by adjusting the $\rm sin \phi$ coefficients (see Appendix B for details).

A dataset of corona brightness from eclipse observations is typically necessary to develop the formulas presented above. In order to achieve the most accurate results, intricate computations and various approaches are necessary. For instance, the outcomes of the method, which are outlined in the following section, were obtained through successive approximations and extensive trials, each time improving the computations slightly. Similarly, another approach, which incorporates van de Hulst's model, was devised by \cite{KH1958}. This article derives the $K$ corona luminance and the corona electron density through the assumption that the polarization arises from $K$ corona light and that $F$ corona light is unpolarized. Consequently, the subsequent equation was formulated as 
\begin{equation}\label{eqn:3}
	K \, P_\nn{K} = P_{\nn{K+F}} (K+F) = K_\nn{t} - K_\nn{r}
\end{equation}
where $K + F$ represents the total corona brightness. $P_\nn{K}$ and $P_{\nn{K+F}}$ refer to the degree of polarisation of the $K$ corona and the total corona, respectively. Furthermore, $K_\nn{t}$ and $K_\nn{r}$ represent the tangential and radial components of the $K$ corona brightness, respectively. Similar complex computations, as in Van de Hulst's method, were carried out during this study. First, the electron density of the corona was determined for the $K_\nn{t} - K_\nn{r}$ component. Subsequently, the $K_\nn{t}$ component was calculated through a reverse calculation of Van de Hulst's equation. Using these values of the $K_\nn{t} - K_\nn{r}$ and $K_\nn{t}$ components, the observational corona brightness $K$ was obtained \cite[for detailed information, refer to article][]{KH1958}.

The new approach presented here has simpler steps compared to the methods mentioned above. The electron density of the corona is computed without time-consuming calculations, using only the luminosity $K$ and the degree of polarization of the corona. As an approximation for the calculations, two new electron densities $N_{\nn{t-r}}$ and $N_\nn{t}$ are defined for the components $K_\nn{t} - K_\nn{r}$ and $K_\nn{t}$, respectively.

Nowadays, as a result of the developing technological possibilities, different methods have been developed to calculate coronal electron densities \citep{BA2020,DZ2023}. However, these methods are quite different from the method described here, in terms of both observation type and electron density calculation technique. In addition, due to the lack of numerical results on the electron density for the equatorial and polar regions in these studies, it was not possible to make a comparison with the results given here.

The van de Hulst approach to determining electron density is concisely outlined in Section 2. A full explanation of about new approximation is given in Section 3. Subsequently, Section 4 presents the validation of the new method utilizing model values from Table 5A in van de Hulst's article. In Section 5, an instance of the new method's application is showcased, featuring the numerical values acquired during the full solar eclipse on the 29th of March, 2006. The Discussion section concludes the article by detailing the benefits and advantages of the novel methodology.

\section{van de Hulst's Method for Calculating the Electron Density}\label{sec:2}
The content of this section is a brief overview of the author's original article, providing only a  basic outline of the method. For more comprehensive information, it is advisable to refer to \cite{VDH1950}. Most of the explanations given here, such as formulae and figures, are also necessary for a better understanding of the new approach presented in the following section.

\begin{figure}[t!]  
	\centerline{\includegraphics[page=1, width=0.55\linewidth]{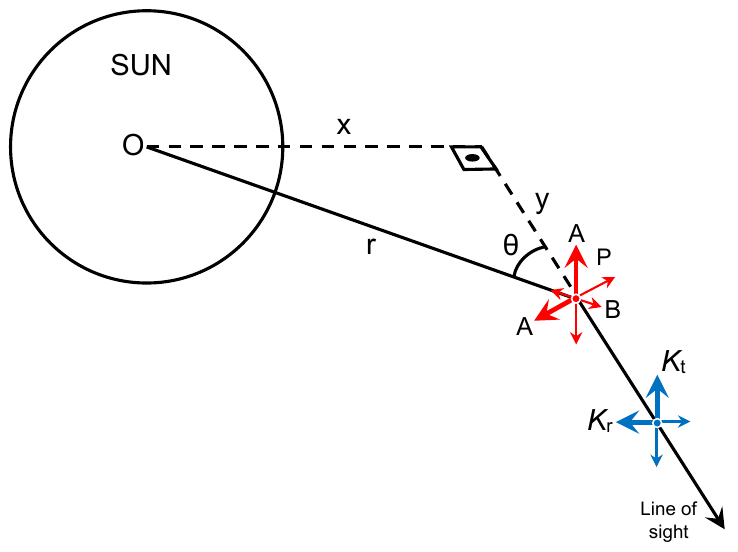}}
	\caption{Geometrical representation of both the scattered light incident on a vibrating electron and its intensity components \citep[reconstructed from][]{VDH1950}.} \label{geo_hulst} 
\end{figure}

The observed brightness of the corona is assumed to be the light scattered by the free electrons \citep{SA1879, BS1937}. Therefore, this brightness should be directly proportional to the density of electrons in the corona. In order to visualise this scenario, consider a single ray of light hitting a vibrating electron (at point P) and reflecting in the direction of the observer (Figure \ref{geo_hulst}). Here, $r$ and $x$ represent the actual distance and projected distance of the light from the centre of the disc, respectively, whilst $\theta$ denotes the angle separating the incoming light direction and the line of sight. Applying the formula presented by \cite{VDH1950}, the total intensity of the light scattered per second per unit solid angle by a column with a cross-section of 1 cm$^2$ is determined using
\begin{equation}\label{eqn:4}
	K(x) = C \int\limits_{x}^{\infty} N(r) \Biggr\{ \bigg( 2 - \frac{x^2}{r^2} \bigg) A(r) + \frac{x^2}{r^2} B(r) \Biggr\} \frac{rdr}{\sqrt{r^2 - x^2}}
\end{equation}
the equations below are expressed in terms of the tangential and radial components of this reflected light,  
\begin{equation}\label{eqn:5}
	\begin{split}
		K_\nn{t} (x) &= C \int\limits_{x}^{\infty} N(r) A(r) \frac{rdr}{\sqrt{r^2 - x^2}} \\[1.5ex]
		K_\nn{t} (x) - K_{\rm r} (x) &= C \int\limits_{x}^{\infty} N(r) \biggr\{ A(r) - B(r) \biggr\} \frac{x^2 dr}{r \sqrt{r^2 - x^2}}
	\end{split}
\end{equation}
where $A$ and $B$ represent the lengths of semi-major and semi-minor axis of the vibration ellipsoid, respectively. The constant C is equal to 3/4 $R_{\odot} \sigma = 3.44 \times 10^{-14} $cm$^3$, where $R_{\odot}\,(= 6.96\times10^{10}\nn{cm})$ represents the solar radius, and $\sigma\,(= 0.66\times10^{-24}\nn{cm}^2$) is the electron scattering cross-section. The primary issue here is to divide the known $K(x)$ intensity into two parts $K_\nn{t}(x)$ and $K_\nn{r}(x)$, and solve the integrals in such a way that both equations produce the same electron density $N(r)$. van de Hulst originally defined the coronal brightness components $K_\nn{t}(r)$ and $K_\nn{t}(r) - K_\nn{r}(r)$ through the use of coronal intensity $K(x)$ and its model polarization degree $p(x)$ as
\begin{align}\label{eqn:6&7}
	K_\nn{t} (x) &= 1/2 [1 + p(x)]K(x) \\[1.2ex]
	K_\nn{t} (x) - K_\nn{r} (x) &= p(x) K(x)
\end{align}
This can also be expressed as a power series in the form of 
\begin{align}\label{eqn:8&9}
	K_\nn{t} (x) &= \sum_s h_s \, x^{-s} \\[1.2ex]
	K_\nn{t} (x) - K_\nn{r} (x) &= \sum_s k_s \, x^{-s}
\end{align}
where $\sum_s h_s \, x^{-s}$ represents with three elements, namely A x$^{-a}$ + B x$^{-b}$ + C x$^{-c}$. van de Hulst then made another approach and assumed that the solution of integrals given in Equation \ref{eqn:5} was of the following form;
\begin{alignat}{2}\label{eqn:10&11}
	r\, C\, N(r)\, A (r)         &= \sum_s \frac{h_s}{a_\nn{-1}} \, r^{-s} && = K_\nn{t} (r) \\[1.2ex]
	r\, C\, N(r) \big\{ A(r) - B(r) \big\} &= \sum_s \frac{k_s}{a_\nn{s+1}} \, r^{-s} && = K_\nn{t} (r) - K_\nn{r} (r) 
\end{alignat}
where 
\begin{equation}\label{eqn:12}
	a_{\rm s} = \int\limits_{0}^{\pi / 2} \sin^n \theta \, d\theta = \frac{\pi}{2^{n+1}} \frac{n!}{\{(n/2)!\}^2}
\end{equation}
The electron densities can now be calculated from Equations 10 and 11. Firstly, the coefficients $h_\nn{s}$, $k_\nn{s}$ and $s$ of Equations 8 and 9 are obtained by making polynomial fit to the calculated values of Equations 6 and 7 separately. Then, the right-hand sides of Equations 10 and 11 are calculated respectively by using a new polynomial function formed with these coefficients. $r$, C, $A(r)$ and $B(r)$ are precomputable values in this approach (refer to \cite{VDH1950}'s article for calculation of these values). At this point, the calculated electron densities $N(r)$ in both Equations 10 and 11 must show the same value. If not, a method of successive approximations is used, replacing both $K_\nn{t}(r)$ and $K_\nn{t}(r) - K_\nn{r}(r)$ by a reduction as small as
\begin{align}\label{eqn:13}
	K_\nn{t} (r) &= \big\{ 1 + \epsilon \, p \big\} K_\nn{t}^{'} \\[1.2ex]
	K_\nn{t} (r) - K_\nn{r} (r) &= \big\{ 1 + \epsilon \, (1 + p) \big\} (K_\nn{t}^{'} - K_\nn{r}^{'})
\end{align}
where $\epsilon$ is a value not exceeding $\pm0.05$. $K_\nn{t}^{'}$ and $K_\nn{t}^{'} - K_\nn{r}^{'}$ are pre-computed values of $K_\nn{t}(r)$ and $K_\nn{t}(r) - K_\nn{r}(r)$. These computations are repeated, altering $\epsilon$ in each iteration, until both $K_\nn{t}(r)$ and $K_\nn{t}(r) - K_\nn{r}(r)$ show the same electron density in Equation 10 and 11. Although van de Hulst has achieved satisfactory results for the electron densities of the model corona using this methodology, its practical application is rather challenging and requires multiple attempts of unspecified numbers.

\section{New Approach for Coronal Electron Density}\label{sec:3}
The values of $K_\nn{t}(x)$ and $K_\nn{t}(x) - K_\nn{r}(x)$ computed in Equations 6 and 7 are numerically different from each other. Thus, this difference should also be valid for Equations 10 and 11. Therefore, the electron density $N(r)$ in each equation must be different as well. At this point, as a novel approach,  the value of $N(r)$ in each equation is given a different nomenclature, defined as
\begin{alignat}{2}\label{eqn:15&16}
	N_\nn{t} (r) &= \frac{1}{r\, \nn{C}}\frac{K_\nn{t}(r)}{A (r)} \\[1.2ex]
	N_{\nn{t-r}}(r) &= \frac{1}{r\, \nn{C}}\frac{K_\nn{t}(r) - K_\nn{r} (r)}{A(r) - B(r)}
    \\[-4ex] \nonumber
\end{alignat}
where $N_\nn{t}(r)$ represents the electron density for $K_\nn{t}(r)$ and $N_{\nn{t-r}}(r)$ represents $K_\nn{t}(r) - K_\nn{r}(r)$. On the other hand, by eliminating the $p(x)K(x)$ values in Equations 6 and 7, the corona intensity $K(x)$ can be expressed in terms of $K_\nn{t}(x)$ and $K_\nn{t}(x) - K_\nn{r}(x)$ as
\begin{equation}\label{eqn:17}
	K_\nn{t} + K_\nn{r} = K = 2 \, K_\nn{t} - (K_\nn{t} - K_\nn{r})
\end{equation}
Considering Equation 4 or Equations 10 and 11, the corona intensity $K(r)$ is linearly proportional to the electron density $N(r)$. Therefore, any valid conclusion drawn between the components of the $K(r)$ corona (Equation \ref{eqn:17}) can also be drawn between the components of the electron density (Equations 15 and 16). Accordingly, the electron density $N$ can be expressed  in a similar manner as
\begin{equation}\label{eqn:18}
	N = 2 \, N_\nn{t}  - N_{\nn{t - r}}
\end{equation}
With this newly derived equation, it is now possible to calculate the electron density for a known $K(x)$ intensity using only the values of the $K_\nn{t}(r)$ and $K_\nn{t}(r) - K_\nn{r}(r)$ components.

\section{Validating the New Approach}\label{sec:4}
The newly developed method was tested using the $K$ corona brightness and the polarization degree values of the van de Hulst model as an observational corona values. The model corona values such as $K_\nn{t} + K_\nn{r}$, $K_\nn{t}$ and $K_\nn{t} - K_\nn{r}$ are taken from Table 5A in \cite{VDH1950}'s article. First, the polynomial coefficients $h_\nn{s}$, $k_\nn{s}$ and $s$ in Equations 8 and 9 were obtained by fitting a separate curve to the brightness values $K_\nn{t}$ and $K_\nn{t} - K_\nn{r}$. Then, new polynomial functions are created using these new coefficients produced by $h_\nn{s} / a_{\nn{s-1}}$ and $k_\nn{s} / a_{\nn{s+1}}$ (shown in Equation 10 and 11). These new functions are referred to as \qq{generated functions} (GFs), and are represented as 
\begin{equation}\label{eqn:19}
	f(K_\nn{t} - K_\nn{r}) = \sum_s \frac{k_\nn{s}}{a_{\nn{s+1}}} \, r^{\nn{-s}}   \quad \text{and} \quad 
	f(K_\nn{t}) = \sum_s \frac{h_\nn{s}}{a_{\nn{s-1}}} \, r^{\nn{-s}}  
\end{equation}
After calculating the electron densities $N_\nn{t}$ and $N_{\nn{t-r}}$ for the brightnesses $K_\nn{t}$ and $K_\nn{t} - K_\nn{r}$ using Equations 15 and 16, the total electron density is determined by combining these values with Equation 18. The results obtained for the equatorial region of van de Hulst model are shown in Table 1. The $K$ corona brightness values of the minimum type model are shown on the left side of the table. Also, the coefficients (A, B, C, a, b and c) of the fitted function for each component are listed under its column at the left-bottom side. For the fitting process, a three-element polynomial is employed, given by (A $r^{\nn{-a}}$ + B $r^{\nn{-b}}$ + C $r^{\nn{-c}}$). Furthermore, values for $a_{\nn{s+1}}$ and $a_{\nn{s-1}}$, calculated using Equation 12 with $a$, $b$ and $c$ coefficients, are listed at bottom of the left side. The computed values of GFs $f(K_\nn{t}$) and $f(K_\nn{t} - K_\nn{r}$) are presented in the first two columns of the right side of Table 1 and the coefficients utilized to construct these functions are listed underneath these values. The calculated electron densities $N_\nn{t}$ and $N_{\nn{t-r}}$, and the total electron density $N$ are exhibited in their corresponding columns on the right side of Table \ref{calc_hulst}.

The electron density calculations were repeated for the polar region of the van de Hulst model, resulting in the same level of agreement.  Table \ref{comp_hulst} presents the electron density values attained by the new method for both the equatorial and polar regions of the minimum type corona, alongside the values obtained by \cite{VDH1950}.  From the table, it can be seen that the new method's values match closely with those of the van de Hulst model for both the equatorial and polar regions. A similar comparison was made using the $K$ corona values and polarization degree values of the \cite{ACW1973}. The same agreement was also achieved for these values. The computed values of the new method for \cite{ACW1973} values are shown in Table \ref{calc_allen} and Table \ref{comp_allen}, respectively.
\newcommand{\ha}{\hspace*{3mm}}
\newcommand{\hb}{\hspace*{2mm}}
\begin{table*}[b!]
	\setlength{\tabcolsep}{3.3mm}
	\setlength\extrarowheight{0.1pt}
	\centering
	\caption{Calculation results for the equatorial region of the minimum-type corona of \protect\cite{VDH1950} computed using the new approach. Brightness is in units of 10$^{-8} I_{\odot}$, and electron density is in units of 10$^6$ cm$^{-3}$.}\label{calc_hulst}
	\tnormalsize
	\begin{tabular}{+l^r^r^r^r|^r^r^r^r^r}
		\cline{1-10}\rowstyle{\boldmath\bfseries}%
		$r$ & $K$\ha & $P_\nn{K}$ & $K_\nn{t}-K_\nn{r}$ & $K_\nn{t}$\ha & $f(K_\text{t}-K_\nn{r})$ & $f(K_\nn{t})$ 
		& $N_{\nn{t-r}}$ & $N_\nn{t}$\ha & $N$\hb \\
		\cline{1-10}  
		\B1.0  & 300.40 & 0.18 & 54.40\hb & 177.40 & 104.03\ha & 470.10 & 239.66 & 226.68 & \B213.7 \\
		\B1.03 & 202.80 & 0.24 & 48.00\hb & 125.40 & 111.69\ha & 328.27 & 171.88 & 176.05 & \B180.2 \\
		\B1.06 & 141.30 & 0.28 & 39.10\hb &  90.20 &  94.66\ha & 230.57 & 128.04 & 130.48 & \B132.9 \\
		\B1.1  &  91.10 & 0.32 & 29.30\hb &  60.20 &  70.99\ha & 148.07 &  87.24 &  88.68 & \B90.1  \\
		\B1.2  &  37.10 & 0.41 & 15.14\hb &  26.12 &  35.53\ha &  58.50 &  39.40 &  39.18 & \B38.9  \\
		\B1.3  &  18.50 & 0.46 &  8.56\hb &  13.53 &  19.72\ha &  28.63 &  21.28 &  21.02 & \B20.8  \\
		\B1.5  &   6.20 & 0.54 &  3.34\hb &   4.77 &   7.46\ha &   9.69 &   8.24 &   8.29 & \B 8.3  \\
		\B1.7  &   2.57 & 0.59 &  1.51\hb &   2.04 &   3.33\ha &   4.06 &   3.90 &   3.96 & \B 4.0  \\
		\B2.0  &   0.85 & 0.62 &  0.53\hb &   0.69 &   1.19\ha &   1.35 &   1.55 &   1.55 & \B 1.5  \\
		\B2.6  &   0.16 & 0.66 &  0.11\hb &   0.13 &   0.23\ha &   0.23 &   0.37 &   0.34 & \B 0.3  \\
		\B3.0  &   0.07 & 0.65 &  0.05\hb &   0.06 &   0.09\ha &   0.09 &   0.17 &   0.15 & \B 0.1  \\
		\B4.0  &   0.02 & 0.61 &  0.01\hb &   0.02 &   0.02\ha &   0.01 &   0.04 &   0.03 & \B 0.02 \\
		\cline{1-10}\rowstyle{\boldmath\bfseries}%
		& & & $k_\nn{s}$\hb & $h_\nn{s}$\ha & $k_\nn{s}$/$a_{\nn{s+1}}$ & $h_\nn{s}$/$a_{\nn{s-1}}$ & & &  \\
		\cline{3-7}
		&  & \B A &  40.50 &  71.98 &  90.03\hb & 143.68 &       &       &  \\
		&  & \B B &  26.66 & 121.28 &  83.22\hb & 398.10 &       &       &  \\
		&  & \B C & -12.77 & -15.86 & -69.22\hb & -71.68 &       &       &  \\
		&  & \B a &   6.25 &   6.74 &   6.25\hb &   6.74 &       &       &  \\
		&  & \B b &  13.80 &  17.42 &  13.80\hb &  17.42 &       &       &  \\
		&  & \B c &  44.63 &  32.58 &  44.63\hb &  32.58 &       &       &  \\[-3.0ex]
		&  &       &       &       &       &       &       &       &  \\
		\cline{3-7}
		\rowstyle{\boldmath\bfseries}%
		&  &       & $a_{\nn{s+1}}$ & $a_{\nn{s-1}}$ &   &    &    &   &  \\
		\cline{4-5}
		&  & for a & 0.4499 & 0.5010 &       &       &       &       &  \\
		&  & for b & 0.3203 & 0.3046 &       &       &       &       &  \\
		&  & for c & 0.1845 & 0.2213 &       &       &       &       &  \\
		\cline{1-10} 
	\end{tabular}
\end{table*}

\newcommand{\ka}{\hspace*{3mm}}
\newcommand{\kb}{\hspace*{2mm}}
\begin{table*}[t!]
	\setlength{\tabcolsep}{3.3mm}
	\setlength\extrarowheight{0.5pt}
	\centering
	\caption{Calculation results for the equatorial region of the minimum-type corona of \protect\cite{ACW1973} computed using the new approach. Brightness is in units of 10$^{-8} I_{\odot}$, and electron density is in units of 10$^6$ cm$^{-3}$.}\label{calc_allen}
	\tnormalsize
	\begin{tabular}{+l^r^r^r^r|^r^r^r^r^r}
		\cline{1-10}\rowstyle{\boldmath\bfseries}%
		$r$ & $K$\ka & $P_\nn{K}$ & $K_\nn{t}-K_\nn{r}$ & $K_\nn{t}$\ka & $f(K_\text{t}-K_\nn{r})$ & $f(K_\nn{t})$ 
		& $N_{\nn{t-r}}$ & $N_\nn{t}$\ka & $N$\kb \\
		\cline{1-10}  
		\B1.01 & 269.15 & 0.22 & 58.41\kb & 163.78 & 175.31\ka & 532.27 & 317.89 & 271.40 & \B224.9 \\
		\B1.03 & 199.53 & 0.23 & 46.09\kb & 122.81 & 115.52\ka & 333.43 & 177.77 & 178.82 & \B179.9 \\
		\B1.06 & 144.54 & 0.25 & 36.28\kb &  90.41 &  86.75\ka & 226.47 & 117.34 & 128.16 & \B139.0 \\
		\B1.1  & 102.33 & 0.28 & 28.24\kb &  65.29 &  65.84\ka & 157.67 &  80.91 &  94.43 & \B107.9 \\
		\B1.2  &  44.67 & 0.33 & 14.74\kb &  29.70 &  35.16\ka &  71.74 &  39.00 &  48.05 & \B 57.1 \\
		\B1.4  &  12.02 & 0.40 &  4.85\kb &   8.43 &  11.59\ka &  18.79 &  12.55 &  14.95 & \B 17.3 \\
		\B1.6  &   4.68 & 0.42 &  1.97\kb &   3.33 &   4.43\ka &   6.43 &   5.03 &   5.89 & \B  6.7 \\
		\B1.8  &   2.00 & 0.39 &  0.77\kb &   1.38 &   1.90\ka &   2.77 &   2.30 &   2.86 & \B  3.4 \\
		\B2.0  &   1.00 & 0.34 &  0.34\kb &   0.67 &   0.89\ka &   1.43 &   1.16 &   1.64 & \B  2.1 \\
		\B2.2  &   0.60 & 0.30 &  0.18\kb &   0.39 &   0.45\ka &   0.85 &   0.62 &   1.07 & \B  1.5 \\
		\B2.5  &   0.27 & 0.26 &  0.07\kb &   0.17 &   0.18\ka &   0.46 &   0.28 &   0.66 & \B  1.0 \\
		\B3.0  &   0.10 & 0.20 &  0.02\kb &   0.06 &   0.05\ka &   0.21 &   0.09 &   0.37 & \B  0.7 \\
		\B4.0  &   0.03 & 0.13 & 0.004\kb &   0.02 &   0.01\ka &   0.07 &   0.01 &   0.17 & \B  0.3 \\
		\cline{1-10}
	\end{tabular}
\end{table*}

\begin{table}[t!]%
	\setlength{\tabcolsep}{10.5pt}
	\setlength{\extrarowheight}{0.8pt}
	\tnormalsize
	\parbox[t][][t]{.49\linewidth}{
		\centering
		\caption{Comparison of electron densities obtained using the new method with those from the \protect\cite{VDH1950} values. Electron density is in units of 10$^6$ cm$^{-3}$.}\label{comp_hulst}
		\setlength{\extrarowheight}{1.28pt}
		\begin{tabular}{+l^r^r|^r^r}
			\cline{1-5}
			(a)& \multicolumn{2}{c|}{van de Hulst $N$} & \multicolumn{2}{c}{This study $N$} \\
			\cline{2-5} \rowstyle{\boldmath\bfseries} $r$ & Equator & Polar & Equator & Polar\\
			\cline{1-5}
			\B1.0  & 227.0 & 174.0 & 213.7 & 170.3 \\
			\B1.03 & 178.0 & 127.0 & 180.2 & 127.0 \\
			\B1.06 & 132.0 &  87.2 & 132.9 &  86.9 \\
			\B1.1  &  90.0 &  53.2 &  90.1 &  52.9 \\
			\B1.2  &  39.8 &  16.3 &  39.0 &  15.8 \\
			\B1.3  &  21.2 &  5.98 &  20.8 &   5.8 \\
			\B1.5  &   8.3 &   1.4 &   8.3 &   1.5 \\
			\B1.7  &   4.0 & 0.542 &   4.0 & 0.620 \\
			\B2.0  & 1.580 & 0.196 & 1.544 & 0.199 \\
			\B2.6  & 0.374 & 0.040 & 0.317 & 0.030 \\
			\B3.0  & 0.176 & 0.017 & 0.131 & 0.010 \\
			\B4.0  & 0.050 & 0.004 & 0.021 & 0.001 \\
			\cline{1-5}
		\end{tabular}%
	}\hfill \parbox[t][][t]{.49\linewidth}{
		\centering
		\caption{Comparison of electron densities obtained using the new method with those from the \protect\cite{ACW1973} values. Electron density is in units of 10$^6$ cm$^{-3}$.}\label{comp_allen}
		\tnormalsize
		\setlength{\extrarowheight}{0.4pt}
		\begin{tabular}{+l^r^r|^r^r}
			\cline{1-5}
			(b)& \multicolumn{2}{c|}{Allen $N$} & \multicolumn{2}{c}{This study $N$} \\
			\cline{2-5} \rowstyle{\boldmath\bfseries} $r$ & Equator & Polar & Equator & Polar \\
			\cline{1-5}
			\B1.01 & 251.2 & 199.5 & 224.9 & 215.4\\
			\B1.03 & 177.8 & 131.8 & 179.9 & 150.1\\
			\B1.06 & 125.9 &  95.5 & 139.0 &  87.7\\
			\B1.1  &  91.2 &  64.6 & 107.9 &  57.6\\
			\B1.2  &  46.8 &  19.9 &  57.1 &  26.2\\
			\B1.4  &  15.1 &   4.4 &  17.3 &   3.4\\
			\B1.6  &   6.8 &   1.3 &   6.7 &   0.5\\
			\B1.8  &   3.6 &   0.6 &   3.4 &   0.1\\
			\B2.0  &   2.0 &   0.3 &   2.1 &  0.06\\
			\B2.2  &   1.3 &   0.2 &   1.5 &  0.04\\
			\B2.5  &   0.6 &   0.1 &   1.0 &  0.02\\
			\B3.0  &   0.3 &  0.05 &   0.7 &  0.01\\
			\B4.0  &   0.1 &  0.02 &   0.3 & 0.001\\
			\cline{1-5}
		\end{tabular}%
	}\vskip 5mm
\end{table}%

\section{Calculated Electron Densities of the 2006 Solar Eclipse}\label{sec:5}
The newly developed method was utilized to compute the electron density of the solar corona as observed during the total eclipse on March 29, 2006, in T\"urkiye. This eclipse observation was carried out with the 8-inch Meade telescope by the staff of the Astronomy and Space Sciences Department of Istanbul University in the Manavgat district of Antalya. During the eclipse event, observations of white light polarization were conducted, and eclipse photographs were taken at three different polarization angles, 0$^\circ$, 60$^\circ$, and 120$^\circ$. A total of 15 photos were taken during totality with an interval of 3$^{\rm m}$ 30$^{\rm s}$ between 11$^{\rm h}$ 55$^{\rm s}$ 10$^{\rm m}$ and 11$^{\rm h}$ 58$^{\rm s}$ 40$^{\rm m}$ UT. Five different exposure times were used in these shoots; 1/2, 1/4, 1/30, 1/60, and 1/125 second. In addition, images of the solar disc were taken at different diaphragm openings before the eclipse for brightness calibration and exposure times used here were the same as those used during the eclipse. After performing brightness calibration and computation of Stokes parameters using polarization images, the total corona brightness ($K+F$) and polarization degree ($P_{\nn{K+F}}$) of the 2006 eclipse obtained by considering the sky with instrumental contribution and active chromospheric regions (see Appendix A for details). 
\begin{figure}[h!]   
	\centerline{\includegraphics[width=0.49\linewidth]{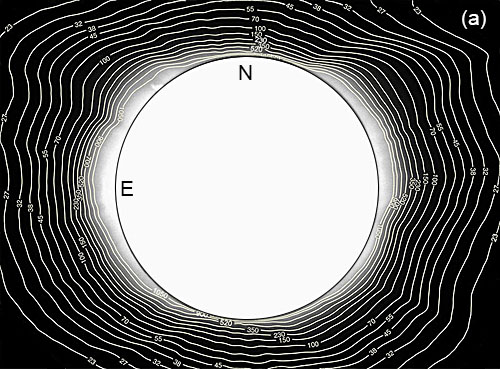}\hspace{1.5mm}
				\includegraphics[width=0.49\linewidth]{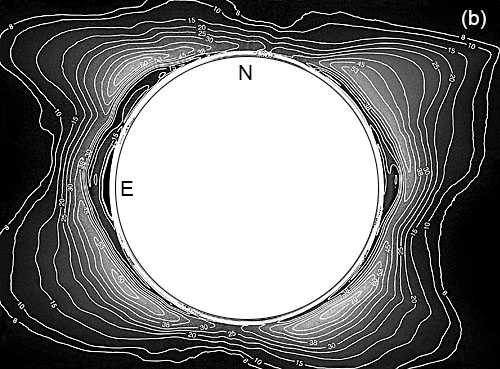}}
	\caption{(a) Isophotes of total corona brightness (values are in units of 10$^{-9} I_{\odot}$) and (b) isolines of polarization degree (values are in percent) of the 29 March 2006 solar eclipse.} \label{ecl2006_obs}
\end{figure}
The isophote plots of total corona brightness and its polarization degree are shown separately in Figure \ref{ecl2006_obs}. The numbers on the isophote lines are in units of 10$^{-9} I_{\odot}$ for the total brightness, and in percent for the polarization degree. The values obtained for both parameters at specific distances from the solar disc are given in Table \ref{ecl2006_res}.

The $K$ corona brightness values are obtained by subtracting $F$ corona model values of van de Hulst (1950) from the observed $K+F$ total brightness values. This process has been carried out with the assumption that $F$ corona does not change much from one solar cycle to another \citep{KK2005, MH2007}. Also, the polarization degree $P_\nn{K}$ of the $K$ corona is determined using the equation of \cite{KH1958}, which is given by
\begin{equation}\label{eqn:20}
	P_{\rm K} = P_{\rm K+F} \Bigg( \frac{K+F}{K} \Bigg).
\end{equation}
For this eclipse, the electron densities in the equatorial and polar regions were calculated using the $K$ corona brightness and its polarization degree. The obtained results are shown in Table \ref{ecl2006_calc}A for the equatorial region and Table \ref{ecl2006_calc}B for the polar region. The calculated electron densities of the equatorial (black circle) and polar (black triangle) regions are shown in Figure \ref{ecl2006_eden} in comparison with the observational values of \cite{NDS1967} and \cite{ACW1973} and model electron density values of \cite{VDH1950} and \cite{SK1970}. It is clear from the figure that the electron density values observed in the 2006 eclipse are in good agreement with comparison values given. Any discrepancies between the observed and model values may be due to the asymmetric brightness distribution caused by the asymmetric distribution of the solar material. This is clearly illustrated in Figure \ref{ecl2006_obs}a, where the equatorial and polar intensity distributions are compared.
\begin{table}[t!]
	\parbox[t][][b]{.35\linewidth}{
		\centerline{\includegraphics[page=2, width=0.98\linewidth]{figures.pdf}}
		\captionof{figure}{Comparison of electron densities in the equatorial and polar regions during the solar eclipse on 29 March 2006 with selected observational data and model values of \protect\cite{VDH1950} and \protect\cite{SK1970}} \label{ecl2006_eden}\vspace{3mm}
	}\hfill\parbox[t][][t]{.63\linewidth}{
		\setlength{\tabcolsep}{9pt}
		\setlength{\extrarowheight}{0.7pt}
		\centering	
		\caption{Observed total corona brightness $K+F$ and polarization degree $P_{\nn{K+F}}$ values of 29 March 2006 eclipse.}\label{ecl2006_res} \tnormalsize
		\begin{tabular}[b]{+c^r^r^r^r^r}
	        \cline{2-3}\cline{5-6}\rowstyle{\boldmath\bfseries}
			& \multicolumn{2}{c}{$K+F$ corona ($\times 10^{-9} I_{\odot}$)} & & \multicolumn{2}{c}{$P_{\nn{K+F}}$ (\%)} \\
			\cline{2-3}\cline{5-6}\rowstyle{\boldmath\bfseries}
			$r$ & Equa. & Polar &    & Equa. & Polar \\
			\cline{1-3}\cline{5-6} 
			\B1.10 & 1311 & 648 &    & 19.0 & 32.4 \\
			\B1.15 & 1046 & 364 &    & 26.7 & 36.3 \\
			\B1.20 &  782 & 201 &    & 36.4 & 33.2 \\
			\B1.25 &  513 & 128 &    & 44.7 & 25.9 \\
			\B1.30 &  293 &  90 &    & 46.1 & 19.6 \\
			\B1.35 &  180 &  70 &    & 39.8 & 15.0 \\
			\B1.40 &  125 &  57 &    & 32.4 & 13.9 \\
			\B1.45 &   94 &  48 &    & 26.8 & 12.3 \\
			\B1.50 &   74 &  42 &    & 22.5 & 11.6 \\
			\B1.55 &   61 &  38 &    & 19.4 & 10.7 \\
			\B1.60 &   52 &  -- &    & 17.1 & --   \\
			\B1.65 &   45 &  -- &    & 15.8 & --   \\
			\B1.70 &   40 &  -- &    & 14.6 & --   \\
			\cline{1-3}\cline{5-6}    
		\end{tabular}%
	}
\end{table}%
\begin{table*}[h!]
\setlength{\tabcolsep}{8pt}
\setlength{\extrarowheight}{0.3pt}
\centering	
\caption{Values used for electron density calculation (\textit{left side}), and results from Eclipse 2006  (\textit{right side}). $K$ values are in units of $10^{-9} I_{\odot}$, and $P$ is in percent.}\label{ecl2006_calc}%
\tnormalsize
\begin{tabular}{+c^r^r^r^r|^r^r^r^r^r}
	\multicolumn{10}{c}{$A -$ equatorial region} \\
	\cline{1-10}\rowstyle{\boldmath\bfseries}
	$r$ & $K$\hb & $P_\nn{K}$ & $K_\nn{t}-K_\nn{r}$ & $K_\nn{t}$\hb
	& $f(K_\nn{t}-K_\nn{r})$ & $f(K_\nn{t})$ & $N_{\nn{t-r}}$ & $N_\nn{t}$\hb & $N_{\nn{equ}}$ \\
	\cline{1-10}  
	\B1.10 & 158.6 & 0.196 & 31.1\hb & 94.8 & 160.5\ha & 269.3 & 197.2 & 161.3 & \B125.4 \\
	\B1.15 &  96.6 & 0.285 & 27.5\hb & 62.0 & 103.4\ha & 171.5 & 118.9 & 109.0 & \B 99.0 \\
	\B1.20 &  72.6 & 0.403 & 28.9\hb & 50.2 &  68.0\ha & 111.4 &  75.4 &  74.6 & \B 73.9 \\
	\B1.25 &  46.0 & 0.517 & 23.8\hb & 34.9 &  45.5\ha &  73.7 &  49.5 &  51.8 & \B 54.1 \\
	\B1.30 &  24.9 & 0.558 & 13.9\hb & 19.4 &  30.9\ha &  49.6 &  33.4 &  36.4 & \B 39.5 \\
	\B1.35 &  14.4 & 0.506 &  7.3\hb & 10.8 &  21.3\ha &  33.9 &  23.0 &  26.0 & \B 28.9 \\
	\B1.40 &   9.4 & 0.433 &  4.1\hb &  6.7 &  14.9\ha &  23.6 &  16.2 &  18.7 & \B 21.3 \\
	\B1.45 &   6.6 & 0.376 &  2.5\hb &  4.6 &  10.6\ha &  16.6 &  11.6 &  13.7 & \B 15.8 \\
	\B1.50 &   5.0 & 0.330 &  1.6\hb &  3.3 &   7.6\ha &  11.8 &   8.4 &  10.1 & \B 11.8 \\
	\B1.55 &   3.9 & 0.298 &  1.2\hb &  2.6 &   5.5\ha &   8.5 &   6.2 &   7.5 & \B  8.9 \\
	\B1.60 &   3.2 & 0.274 &  0.9\hb &  2.1 &   4.0\ha &   6.2 &   4.6 &   5.7 & \B  6.8 \\
	\B1.65 &   2.4 & 0.263 &  0.6\hb &  1.5 &   3.0\ha &   4.6 &   3.5 &   4.3 & \B  5.2 \\
	\B1.70 &   1.9 & 0.249 &  0.5\hb &  1.2 &   2.2\ha &   3.4 &   2.6 &   3.3 & \B  4.0 \\
	\cline{1-10}
	\multicolumn{10}{c}{} \\[-1ex]
	\multicolumn{10}{c}{$B -$ polar region} \\
	\cline{1-10}\rowstyle{\boldmath\bfseries}
	$r$ & $K$\hb & $P_\nn{K}$ & $K_\nn{t}-K_\nn{r}$ & $K_\nn{t}$\hb
	& $f(K_\nn{t}-K_\nn{r})$ & $f(K_\nn{t})$ & $N_{\nn{t-r}}$ & $N_\nn{t}$\hb & $N_{\nn{pol}}$ \\
	\cline{1-10}  
	\B1.10 & 54.8 & 0.395 & 21.6\hb & 38.2 & 76.5\ha & 115.4 & 94.0 & 69.1 & \B44.2 \\
	\B1.15 & 28.5 & 0.425 & 12.1\hb & 20.3 & 36.0\ha &  58.0 & 41.4 & 36.9 & \B32.4 \\
	\B1.20 & 13.7 & 0.382 &  5.2\hb &  9.4 & 18.3\ha &  30.6 & 20.3 & 20.5 & \B20.7 \\
	\B1.25 &  7.6 & 0.316 &  2.4\hb &  5.0 & 10.0\ha &  16.9 & 10.9 & 11.9 & \B12.8 \\
	\B1.30 &  4.7 & 0.275 &  1.3\hb &  3.0 &  5.9\ha &   9.8 &  6.4 &  7.2 & \B 8.0 \\
	\B1.35 &  3.3 & 0.252 &  0.8\hb &  2.1 &  3.7\ha &   5.9 &  4.0 &  4.5 & \B 5.1 \\
	\B1.40 &  2.5 & 0.240 &  0.6\hb &  1.6 &  2.4\ha &   3.7 &  2.6 &  3.0 & \B 3.3 \\
	\B1.45 &  2.1 & 0.258 &  0.5\hb &  1.3 &  1.7\ha &   2.5 &  1.8 &  2.0 & \B 2.2 \\
	\B1.50 &  1.0 & 0.287 &  0.3\hb &  0.7 &  1.2\ha &   1.7 &  1.3 &  1.4 & \B 1.6 \\
	\B1.55 &  0.5 & 0.266 &  0.1\hb &  0.3 &  0.9\ha &   1.2 &  1.0 &  1.0 & \B 1.1 \\
	\B1.60 &  0.3 & 0.264 &  0.1\hb &  0.2 &  0.6\ha &   0.9 &  0.7 &  0.8 & \B 0.8 \\
	\cline{1-10}  
\end{tabular}
\end{table*}

\section{Discussion}\label{sec:6}
The main challenge of the newly developed method is to determine the optimal coefficients ($h_\nn{s}, k_\nn{s}, s $) of the power function in transition from Equations 8--9 to Equations 10--11. This involves a demanding phase of performing numerous fitting curve tests to determine the appropriate coefficients of the three- or two-element power function. Using these coefficients, the observational values of $K_\nn{t}(x)$ and $K_\nn{t}(x) - K_\nn{r}(x)$, which depend on the projection distance, are converted into the values of $K_\nn{t}(r)$ and $K_\nn{t}(r) - K_\nn{r}(r)$, which depend on the true distance.
\begin{figure*}[t!]  
\centerline{\includegraphics[width=0.7\linewidth]{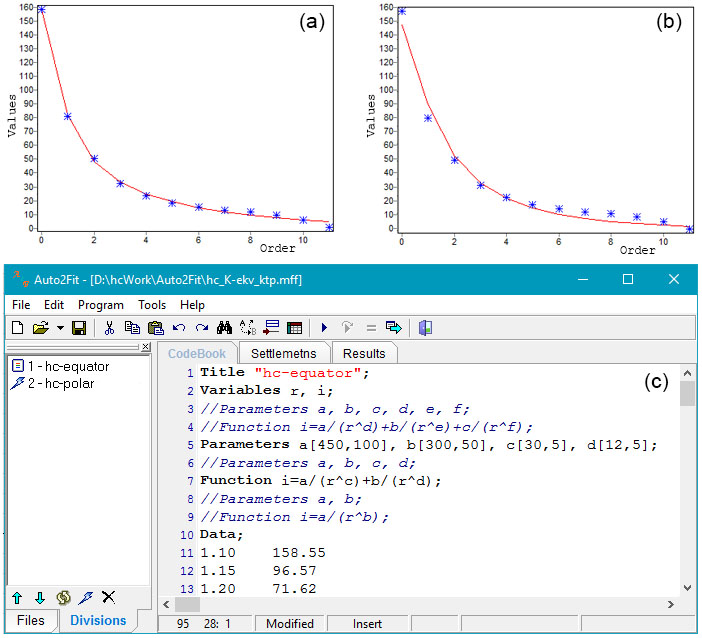}}
\caption{(a) Power function curve fitted without any constraints, (b) power function curve fitted with delimited coefficients, (c) screenshot of a program showing parameters utilized to fit the power function curve under different configurations.} \label{calc_adjust} 
\end{figure*}

During the fitting process, it is crucial to ensure that the fitted curve passes through the overall distribution of the observation points. Attempting to fit the curve close to every observation point is generally ineffective (see Figure \ref{calc_adjust}a) due to inevitable observational errors that cause scattering in the values. Thus, a solution should be devised for the general trend of these observation points (see Figure \ref{calc_adjust}b). There are two possible techniques to accomplish this task. The first option involves fitting a two-element power function curve to the values of a curve obtained by fitting a polynomial equation with one element or six or more elements. The second option involves fitting a power function curve with restricted parameters where each coefficient has a boundary between specified values (refer to Figure \ref{calc_adjust}c). The second method, which is preferred in this study, provides a straightforward and fulfilling solution without requiring additional experimentation. However,  it can be challenging to adjust the limits for limiting coefficients in a consistent manner across different eclipse data. Nevertheless, this is a common occurrence, as each eclipse typically has a unique distribution with its own distinct characteristics. Once the coefficients for the power function are determined for the observational values, generating GFs and obtaining the electron density become straightforward steps in this method.

\begin{figure*}[t!]  
	\centerline{\includegraphics[page=3, width=0.7\linewidth]{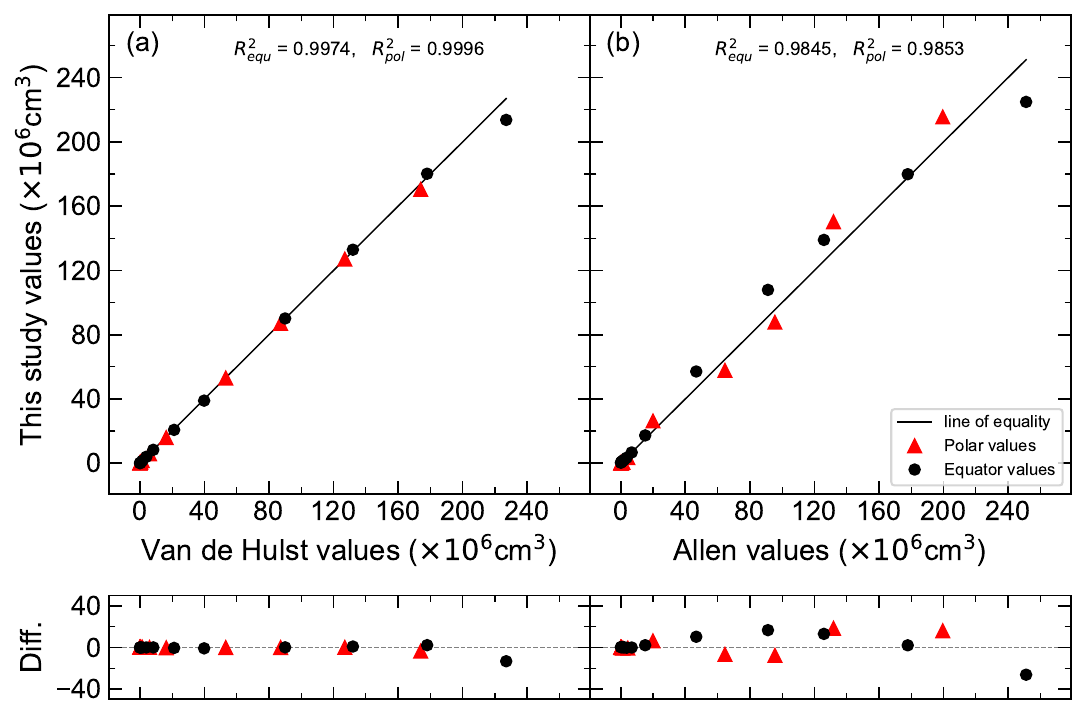}}
	\caption{Comparison of electron density values from (a) van de Hulst's model and (b) Allen's values with those derived from this study. The plots below illustrate the numerical disparity between these two approaches.} \label{calc_comp} 
\end{figure*}

When examining the test results of the electron density for \cite{VDH1950} obtained by the newly introduced method shown in Table \ref{comp_hulst}, it is evident that the compared values are in very good agreement. This fact is more apparent in Figure \ref{calc_comp}a, which confirms the accuracy level of the new method. The similar agreement is also seen for the values of \cite{ACW1973} given in Tables \ref{calc_allen} and \ref{comp_allen}, respectively. As highlighted in Figure \ref{calc_comp}, the electron density values of both the model and the novel method are mainly distributed along the line. When checked for compatibility in individual regions, the coefficient of determination $R^2$ is 0.9974 for the equatorial region and 0.9996 for the polar region of \cite{VDH1950}, while 0.9845 for the equatorial region and 0.9853 for the polar region of \cite{ACW1973}.
Although the results from Eclipse 2006 are quite satisfactory, It would be better to retest this new method with the data from other eclipse observations, especially with the other eclipse results from other researchers who have their own observational electron density data. Thus, the accuracy of the new method will be confirmed by finding similar or conclusive results for these data. For example, when the new method was tested with van de Hulst model values in Section 4, satisfactory electron density results were obtained. Thus, the probability of finding similar consistent results with other observational data is quite high. This should be examined, particularly by other researchers who have measured electron densities using their methods. In order to check the new approach with different observational values, it is intended to contact more than one researcher investigating this topic in the future.

\vspace{8pt}
\begin{description}
    \item[Peer Review:] Externally peer-reviewed.
    \item[Author Contribution:] Conception/Design of study - H.\c{C}.;~ Data Analysis/Interpretation - H.\c{C}.;~Drafting Manuscript - H.\c{C}.;~Critical Revision of Manuscript - H.\c{C}.;~Final Approval and Accountability - H.\c{C}.
    \item[Conflict of Interest:] Authors declared no conflict of interest.
    \item[Financial Disclosure:] Authors declared no financial support.
\end{description}

\section*{Acknowledgements}
Thanks to especially every staff who took part in the 2006 solar eclipse observation. Thanks also to the anonymous referees for their valuable suggestions and comments that improved the manuscript. This work was supported by the Istanbul University Scientific Research Projects Commission with project numbers 24242 and 470/27122005.

\bibliographystyle{mnras}
\bibliography{hcreferences}

\appendix
\section{Calculation procedures of the K corona brightness}
\begin{figure}[h!]
\centerline{\includegraphics[page=4, width=0.65\linewidth]{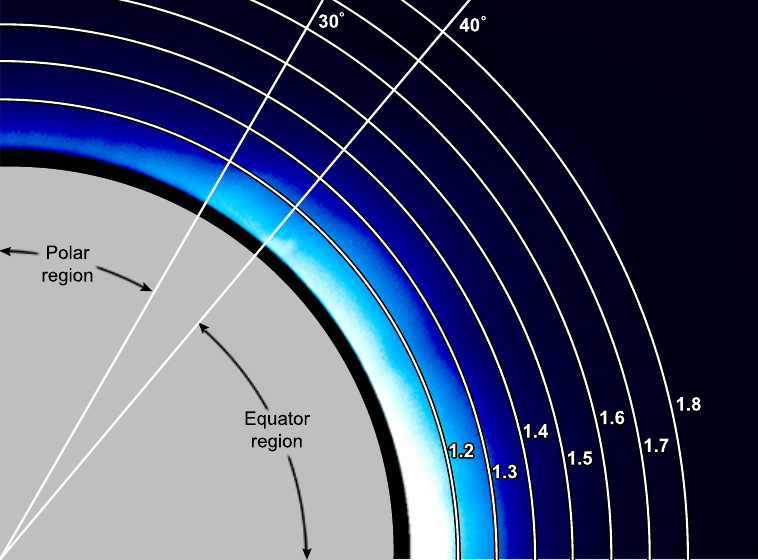}}
\caption{The latitude ranges for the polar and equatorial regions used in a single quadrant. Distances to the solar disc are given in units of solar radius.}
\label{cor_range}
\end{figure}

First of all, as a reminder, all procedures relevant to this section are given comprehensively in \cite{HC2017}'s article. Please review this article for more information. A very brief summary of the general steps of this procedure is given verbally here. In order to determine the brightness of the $K$ corona, a brightness calibration must first be performed. This requires taking images at different diaphragm openings with the same exposure times as used for polarized images. Once the intensity calibration function \citep[defined in][]{HC2017} has been obtained by using these solar disc images, the brightness of the corona in all polarized images is calculated by normalizing their intensity. Then, the average corona brightness values are determined for the polar region between latitudes 0$^\circ$--30$^\circ$ and for the equatorial region between latitudes 40$^\circ$--90$^\circ$, separately (Figure \ref{cor_range}). At this stage, regions with chromospheric structure were excluded from the calculation, taking into account their distance range from the Sun's surface, to avoid erroneous increases in brightness.

\section{Explanations about latitude-dependent coronal brightness calculation}
\begin{figure}[h!]
\centerline{\includegraphics[page=5, width=0.6\linewidth]{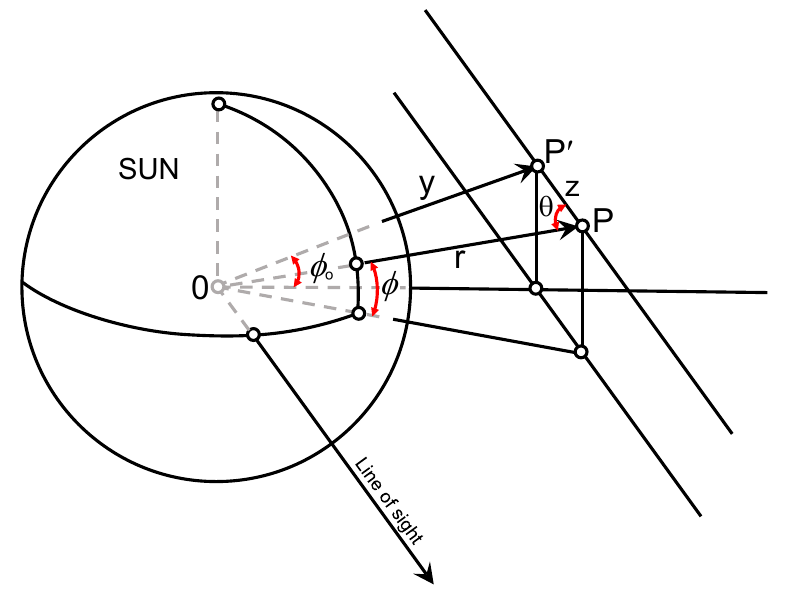}}
\caption{Latitude-dependent diagram of an arbitrary point in the solar corona \protect\cite[reconstructed from][]{SK1970}.}
\label{saito_graph}
\end{figure}

A very brief summary of the explanations on this subject from the article of \cite{SK1970} is given here. Please refer to that article for more information. To describe the brightness of an arbitrary point in the solar corona, the graphical situation shown in Figure \ref{saito_graph} is considered. In this figure, $P$ is the point in question, $P'$ is the projection of $P$ on the celestial plane, $\phi$ is the heliographic latitude of $P$, $\phi_0$ is the projected angle of on the celestial plane, $\theta$ is angle $OPP'$ and $z$ is the line-of-sight length.

\noindent In that paper, the author obtained the following equations from Equation \ref{eqn:5} 
\begin{equation}\label{elat}
	\begin{matrix}
		I_\nn{t} - I_\nn{r} = C\ 2y \int\limits_{0}^{\pi/2} N(r, \phiup) (A-B)\nn{d}\thetaup \\[0.5ex]
		I_\nn{t} = C\ 2y \int\limits_{0}^{\pi/2} N(r, \phiup) (A)\frac{\nn{d}\thetaup}{\rm \sin^2\thetaup}
	\end{matrix}
\end{equation}
under certain assumptions with
\begin{equation*}
\begin{matrix}
	\sin\phiup = \sin\phiup_0 \,\sin\thetaup\,,\quad r\,\sin\thetaup = y\\[1.5ex]
	N = \frac{N_0}{R^n} = N_0\,k^n\,\sin^n\,\thetaup\,,\quad
	N(r,\phi) = \sum N_{0,i}\,\frac{1-f_i\,\sin^{si}\,\phiup}{r^{ni}}
\end{matrix}
\end{equation*}
where $k$ is the modulus of the integrals and a number between 0 and 1, $n$ is a real number, $f_i$ and $s_i$ are positive real number. As a result of arranging the Equation \ref{elat}, the following formula is obtained for latitude-dependent $K$ corona brightness.

\begin{equation}
\begin{split}
(I_\nn{t} \pm I_\nn{r})_{\phiup_0} = (I_\nn{t} \pm I_\nn{r})_\nn{equ}
	&- \frac{\sin\phiup_0}{k}\,0.5\times 5.365\times 10^{-6}(I_\nn{t} \pm I_\nn{r})_{17}\\
	&\hspace*{-4mm}+ \frac{\sin\phiup_0}{k}\,0.95\times 2.752\times 10^{-6}(I_\nn{t} \pm I_\nn{r})_{7}\\
	&\hspace*{-4mm}+ \frac{\sin^{0.5}\phiup_0}{k^{0.5}}\,1.0\times 0.0436\times 10^{-6}(I_\nn{t} \pm I_\nn{r})_{3}
\end{split}
\end{equation}

\vspace*{3mm}

\noindent This equation can be expressed as
\begin{align}\label{eden}
N_\nn{e}(r, \phiup) &= \frac{3.09\times 10^8}{r^{16}} \big( 1-{\color{red}0.5}\ \rm sin \phiup \big) \nonumber\\
&+ \frac{1.58\times 10^8}{\mathit{r}^6} \big( 1-{\color{red}0.95}\ sin \phiup \big)\\
&+ \frac{0.0251\times 10^8}{\mathit{r}^{2.5}} \big( 1-\rm {\color{red}1.0}\ sin^{0.5} \phiup \big)\nonumber
\end{align}

\begin{figure}[h!]
\centerline{\includegraphics[page=6, width=0.68\linewidth]{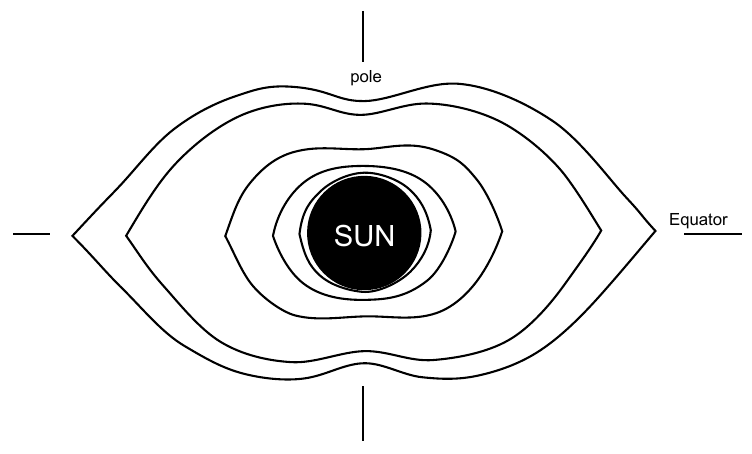}}
\caption{Iso-density curves of electrons in the $K$ corona computed with the Equation \ref{eden} \protect\cite[reconstructed from][]{SK1970}.}
\label{saito_kcor}
\end{figure}

\noindent Using this last equation, the coronal electron density can be visualized as shown Figure \ref{saito_kcor}. As can be seen from Equation \ref{eden}, different profiles for the coronal electron density can be produced by changing the coefficients of $\rm sin\phiup$'s shown in red.

\label{lastpage}
\end{document}